# An Equation for Predicting Binding Strengths of Metal Cations to Protein of Human Serum Transferrin

**Huifang Xu[1 *] and Yifeng Wang[2]**

[1] NASA Astrobiology Institute, Department of Geoscience,

University of Wisconsin - Madison, 1215 W Dayton Street, Madison WI 53706, USA

[2] Sandia National Laboratories, Albuquerque, NM 87185, USA

* Corresponding author: Prof. Huifang Xu

Tel: 608-265-5887 (O); E-mail: hfxu@geology.wisc.edu

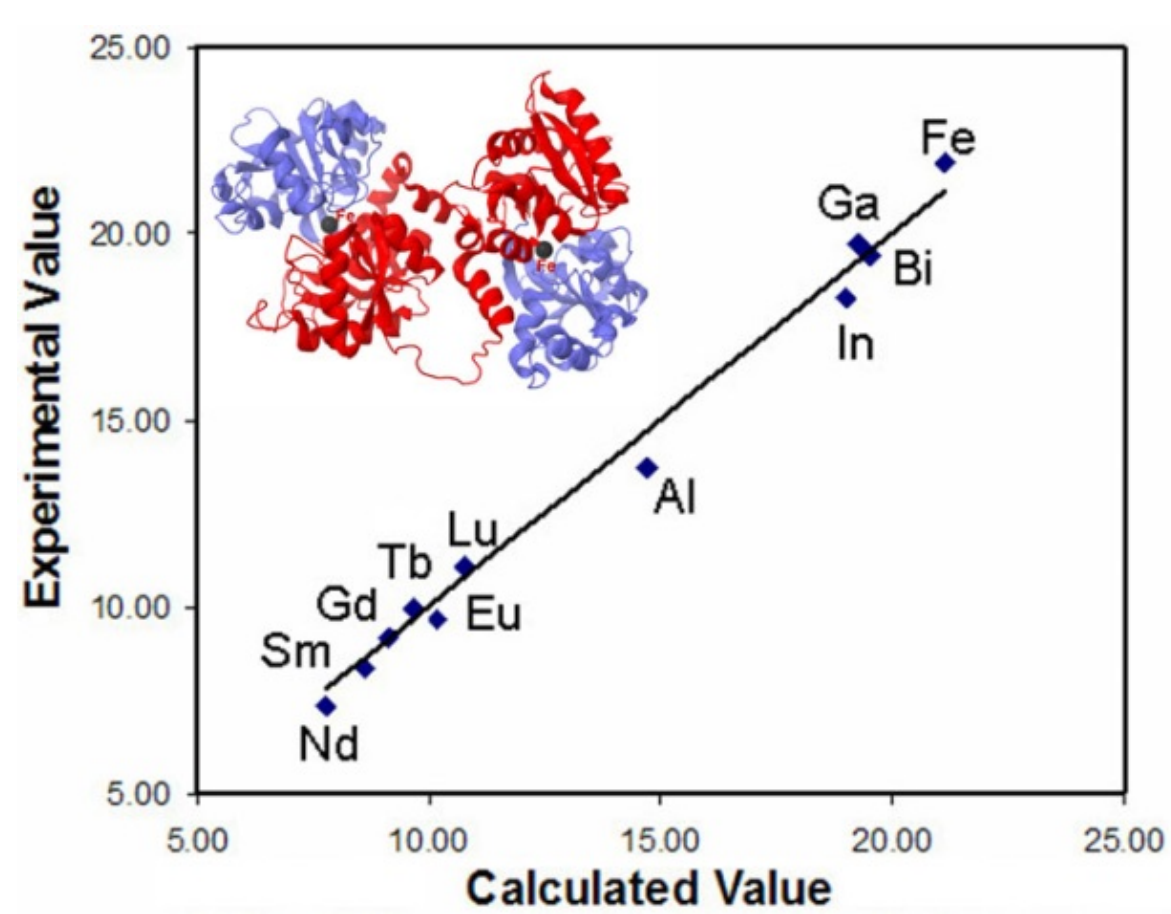

## Abstract

**Because human serum transferrin (hTF) exists freely in serum, it is a potential target for cancer treatment drugs and in curing iron-overloaded conditions in patients via long-term transfusion therapy. The understanding of the interactions between hTF and metal ions is very important for biological, pharmalogical, toxicological, and other protein engineering purposes. In this paper, a simple linear free energy correlation is proposed to predict the binding strength between hTF protein and metal cations. The stability constants ($K_{ML}$) for a family of metal—hTF complexes can be correlated to the non-solvation energies ($\Delta G^0_{n,M^{n+}}$) and the radii ($r_{M^{n+}}$) of cations by equation:**

$$2.303RT\,log K_{ML} = -a^*_{ML}\,\Delta G^0_{n,M^{n+}} - b^{**}_{ML} - \beta^*_{ML}\,r_{M^{n+}} + \Delta G^0_{f,M^{n+}},$$

**where the coefficients $a^*_{ML}$, $b^{**}_{ML}$, and $\beta^*_{ML}$ characterize a particular family of metal-protein complexes. The binding strength is determined by both the physical properties (charge and size or ionic radius, $r_{M^{n+}}$) and chemical properties (non-** **solvation energy, $\Delta G^0_{n,\ M^{n+}}$) of a given cation. The binding strengths of either divalent or and trivalent metals can then be predicted systematically. The predicted stability constants of $Pu^{3+}$—hTF, $Am^{3+}$—hTF, and $Cm^{3+}$—hTF complexes are much lower than that of $Fe^{3+}$—hTF complex. The predicted stability constants of $Co^{3+}$—hTF, $Tl^{3+}$—hTF, $Au^{3+}$—hTF, and $Ru^{3+}$—hTF complexes are higher than that of the $Fe^{3+}$—hTF complex.**

## Introduction

Human serum transferrin (hTF) is a single-chain glycoprotein that transports iron (Fe) in blood [1-4]. Human serum transferrin has a very strong binding force to bind metal ions, such as $Fe^{3+}$, $Ga^{3+}$, and $Al^{3+}$. The hTF molecule contains about 700 amino acids with molecular weight of about 80 kDa [2]. The $M^{3+}$ cation binds with two Tyr, one His, one Asp, and one bidendate $CO_3$[2-5, 6 7-11 12]. The coordination environment of the bonded metals is either an octahedron or a distorted octahedron [4, 6 7-11]. The anion $CO_3^{2-}$ is called a synergistic anion (Figure 1). The $M^{3+}$ cations cannot bind strongly without the synergistic anion [2, 5, 6 7-11]. Although the protein hTF is very complex and looks like poly-dendate ligands, it may be considered as a single hexadendate ligand, such as EDTA or ethylenebis (*o*-hydroxyphenyglycine). The binding strength between metal ion and protein can be characterized by the stability constant of a metal-protein complex [5, 13, 14 15-17]. The stability constants according to the reference are expressed as [5, 13, 17]:

$K_1 = [M^{3+}—hTF]/[M^{3+}][hTF]$,

and

$K_2 = [M^{3+}—(hTF-M^{3+})]/[M^{3+}][M^{3+}- hTF]$.

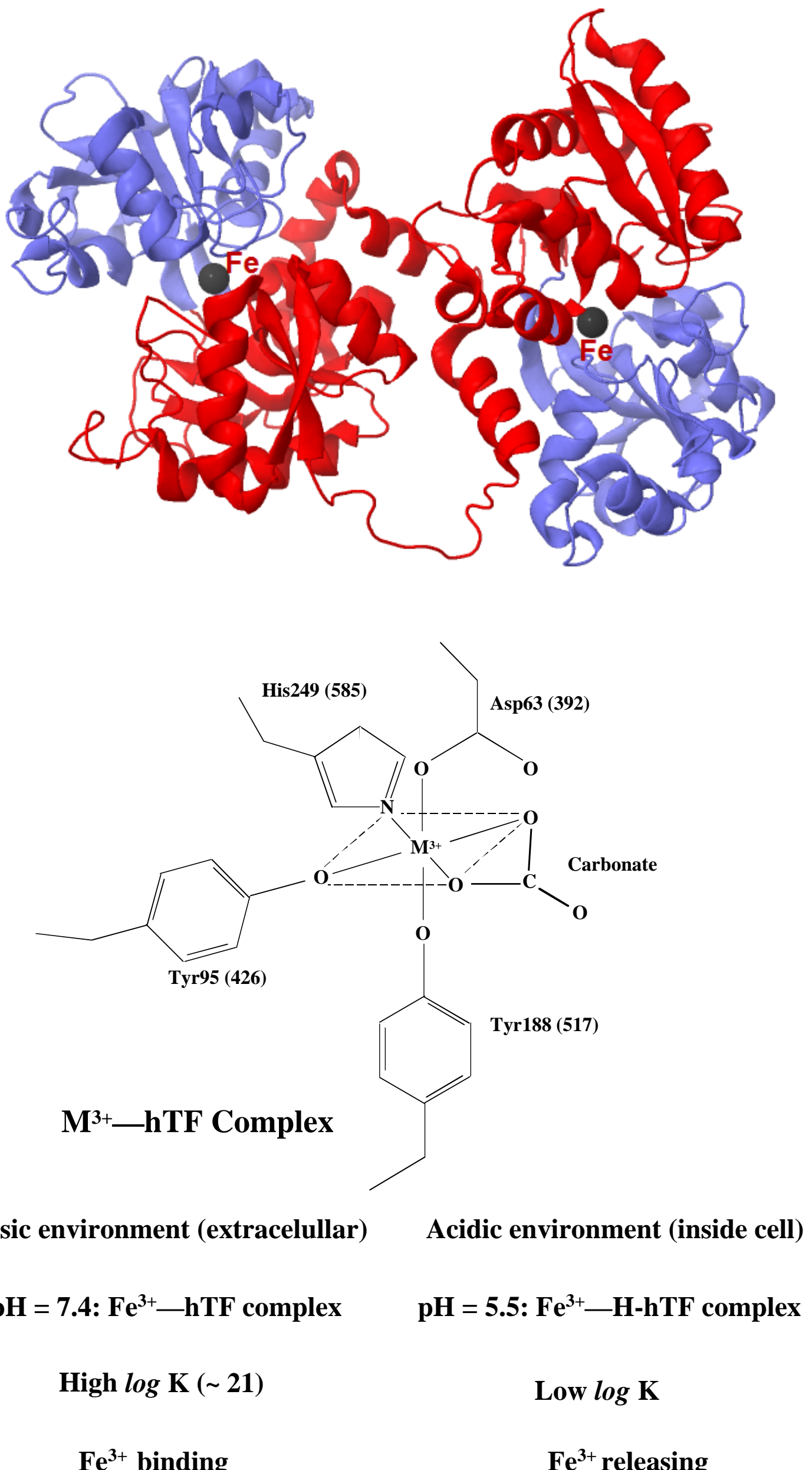

Figure 1: Diagrams showing protein structure of hTF (upper) drawn by using ChemTube3D online program and the coordination environment of $M^{3+}$ cations in the $M^{3+}$—hTF complex (modified from references [6]).

Because human serum transferrin is only 30% saturated with $Fe^{3+}$, the hTF has a high capacity to bind other metals [2, 18, 17]. Therefore, it is important to understand the interaction between hTF and other metals, although it is very difficult to measure the stability constants for metal-hTF complexes. For instance, platinum and gold complexes are used as anticancer and anti-tumor agents [19-23], organometallic ruthenium Complexes are considered as a potential anti-cancer drug [24-28] [29] [30], and the metal ion $Bi^{3+}$ is widely used for variety of gastrointestinal disorders including diarrhea, constipation, gastritis, and ulcers [31, 32] [33] [34] [35] [36] [37] [38]. There are limited number of stability constants for metal-hTF complexes and other metal-protein complexes [5] [17, 39] [40]. Binding constants for some redox-sensitive metals (e.g., $Fe^{2+}$, $Mn^{3+}$ and $Co^{3+}$) are difficult to measure experimentally [41, 42]. Also, metals like Cm, Am, U and others metals associated with nuclear energy and weapons, are hazardous to human health, and it is imperative to be able to predict the transferring binding constants for these metals. These metals, however, are risky to perform experiments upon because of their transuranium radiation dangers) [2, 43]. The existing correlation methods for predicting the stability constants are generally based on the stability constants of metal complexes with small ligands or the hydrolysis constants [14] [16], which are generally not fully available, especially for metal cations like such as $Pu^{3+}$, and $Am^{3+}$. In addition, the methods based on small ligand complexes could lead to large uncertainties between experimental values and predicted values, which could be as large as two to three *log* units for $Fe^{3+}$, $Bi^{3+}$, $Cu^{2+}$, $Gd^{3+}$, $In^{3+}$, and $Al^{3+}$ [14]. Therefore, there is a need for developing a tool that can give a reliable prediction of unknown stability constants that can also be based on the limited number of the existing measurements. Furthermore, as even more thermodynamic data for metal-protein complexes are accumulated, such a tool is also needed for the systematic evaluation of the quality of data collected from multiple sources to develop an internally consistent reliable data set for modeling metal-protein interactions.

**Proposed equation**

Linear free energy relationships have been used for correlating organic reactions (Hammett equation) [44], stabilities of crystalline solids [45] [46], and trace metals bindings at mineral-water interfaces [47]. A similar relationship can be also used for metal—protein binding because the coordination environment for a metal in transferin is much like the polyhedral coordination of metal cations in crystal structures [5, 6] [7-11]. A family of metal-proteins or metal-organic chelate refers to the complexes formed with different metals with a given protein or organic ligand. In each family, metal cations have the same charge. A family of metal complexes can be represented by ($M^{n+}L$), where $M^{n+}$ is a cation with a charge +n, and L represents the ligand of the metal complex (e.g., in $M^{3+}$—hTF metal-protein complexes, the trivalent cations $M^{3+}$ are Fe, Ga, Tl, etc., and L is hTF). The local coordination environment of metal cations in hTF is very similar to that of crystalline minerals like calcite. Because it is analogous to the crystalline structure of solids [45 46-48], the linear free energy correlation is expressed as:

$$\Delta G^0_{f,\,ML} = a^*_{ML}\,\Delta G^0_{n,\,M^{n+}} + b^*_{ML} - \beta^*_{ML}\,r_{M^{n+}}, \qquad (1a)$$

where, $\Delta G^0_{f,\,ML}$ is the Gibbs free energy of formation of the complex ML. The stability of most metal complexes (metalloproteins), for a metal complexation reaction of $M^{n+} + L = M^{n+}L$, is expressed as a stability constant given by [47]:

$$-2.303RT\,log K_{ML} = a^*_{ML}\,\Delta G^0_{n,\,M^{n+}} + b^*_{ML} + \beta^*_{ML}\,r_{M^{n+}} - \Delta G^0_{f,\,M^{n+}} - \Delta G^0_{f,\,L} \qquad (1b)$$

or,

$$-2.303RT\,log K_{ML} = a^*_{ML}\,\Delta G^0_{n,\,M^{n+}} + b^{**}_{ML} + \beta^*_{ML}\,r_{M^{n+}} - \Delta G^0_{f,\,M^{n+}} \qquad (1c)$$

or,

$$log K_{ML} = -(a^*_{ML}\,\Delta G^0_{n,\,M^{n+}} + b^{**}_{ML} + \beta^*_{ML}\,r_{M^{n+}} - \Delta G^0_{f,\,M^{n+}}) / 2.303RT \qquad (1d)$$

where,

$$b^{**}_{ML} = b^*_{ML} - \Delta G^0_{f,\,L}\,. \qquad (1e)$$

In above equations, $\Delta G^0_{f,\,L}$ is the Gibbs free energy of formation of a ligand L or, hTF, $K_{ML}$ is the stability constants of a metal complexation reaction $M^{n+} + L = M^{n+}L$, and the coefficients $a^*_{ML}$, $b^{**}_{ML}$, and $\beta^*_{ML}$ characterize a particular family of metal complexes $M^{n+}L$, which can be calculated by fitting equation (1c) to a limited number of experimentally determined *log* K values; $r_{M^{n+}}$ is the ionic radius of the $M^{n+}$ cation referring to six-fold coordination for divalent [49], trivalent and tetravalent cations [50, 51]; $K_{ML}$ is thermodynamic stability constant of a metal complex, and $\Delta G^0_{n,\,M^{n+}}$ is the standard non-solvation energy, corrected for cation radius [51]. Equation (1d) allows the prediction of the stability constant of a metal complex from the known thermodynamic properties of the corresponding metal cation.

The non-solvation energy $\Delta G^0_{n,\ M^{n+}}$ that corresponds to the unhydrated cation energy, can be calculated by:

$$\Delta G^0_{n,\ M^{n+}} = \Delta G^0_{f,\ M^{n+}} - \Delta G^0_{s,\ M^{n+}}, \qquad (2)$$

where $\Delta G^0_{f,\ M^{n+}}$ and $\Delta G^0_{s,\ M^{n+}}$ represent the standard Gibbs free energy of formation and the standard solvation energy of a bare metal cation respectively. $\Delta G^0_{s,\ M^{n+}}$ can be calculated from the conventional Born solvation coefficients for aqueous cations [45] according to the equation:

$$\Delta G^0_{s,\ M^{n+}} = \omega_{M^{n+}} (1/\varepsilon - 1). \qquad (3)$$

In equation (3), $\varepsilon$ is dielectric constant of water (78.47 at 25 °C). The parameter $\omega_{M^{n+}}$ is the Born solvation coefficients for aqueous cations and can be calculated by:

$$\omega_{M^{n+}} = \omega^{abs}_{M^{n+}} - n\omega^{abs}_{H^+}. \qquad (4)$$

In equation (4), $\omega^{abs}_{H^+}$ is the absolute Born solvation coefficient of $H^+$ (53.87 kcal/mole), and $\omega^{abs}_{M^{n+}}$ is the absolute Born solvation coefficients of the cations with the effective electrostatic radii of aqueous ions ($r_{e,\ M^{n+}}$). They can be calculated by

$$\omega^{abs}_{M^{n+}} = 166.027\ n^2/(\ r_{e,\ M^{n+}}), \qquad (5)$$

$$r_{e,\ M^{n+}} = r_{M^{n+}} + n\ 0.94. \qquad (6)$$

The radii and non-solvation energies of trivalent and divalent cations [49, 50] are listed in Table 1 and Table 2. The Gibbs free energies of metal cations ($\Delta G^0_{f,\ M^{n+}}$) basically increase with decreasing hardness of Pearson's Lewis acids. We may define cations with positive $\Delta G^0_{f,\ M^{n+}}$ to be soft acids, and those with negative $\Delta G^0_{f,\ M^{n+}}$ to be hard acids as discussed below.

## Results and Discussions

All the data used for regression analysis are effective binding constants that have been corrected for bicarbonate concentration by the equation [40, 52]:

$$\mathit{log}K = \mathit{log}K^* - \log\alpha \qquad (7)$$

where,

$$\alpha = K_c\,[HCO_3^-] / (1 + K_c\,[HCO_3^-]). \qquad (8)$$

The value of $K_c$ for the both binding sites is $10^{2.5}$ [52].

Equation (1c) has been applied to $M^{3+}$—hTF (*log* $K_1$) and $M^{3+}$—($M^{3+}$-hTF) (*log* $K_2$) complexes as well (Table 1). The coefficients of $a^*_{ML}$, $\beta^*_{ML}$, and $b^{**}_{ML}$ can be obtained by regression analysis using experimental data measured at a neutral pH condition (see Table 1 for detail). Standard errors for *log* $K_1$ and *log* $K_2$ values are ±0.54 and ±0.76, respectively. Standard error for a well-studied ligand NTA is ±0.36 (Table 1). The discrepancies between measured and calculated values are generally within 0.6 *log* units (Figure 2). Only $Sc^{3+}$ data was not used for regression analysis, because of anomalously large discrepancies (2-3 log units) between the measured and the calculated values (Table 1). There are two sets of data for Al. The data by Harris and Sheldon [52] are very close the calculated values (Table 1). The *log* $K_1$ value by Martin et al is off by ~2 *log* units [53]. There are multiple sets of reported data for Tb (11.2, 7.61; 10.96, 8.52) [40]. The most recent published values (**9.96**, **6.37**) listed in Table 1 are used for regression analysis [54]. It may be necessary to re-determine the binding constants for $Sc^{3+}$, because there are large discrepancies among experimentally measured values similar to Tb and Al [40] [17].

*Table 1: Ionic radii, thermodynamic data for trivalent cations, and stability constants for three families of metal complexes.*

***Insert Table 1 Here***

*Note: hTF = human serum transferring; NTA = Nitrilotriacetic acid ($C_6H_9O_6N$)*

*Note: Radii of the trivalent cations are from reference [50]. The values of $\Delta G_f$ of the cations are from references [51, 55] except for $Bi^{3+}$ from reference [56], $Ti^{3+}$ from reference [57], and $Pu^{3+}$, $Np^{3+}$, and $Am^{3+}$ from references [58] [59]. The log K values of Fe—hTF are from references [13]; the values of Bi—hTF, Ga—hTF, and In—hTF are from references [15, 60] [61] [62]; the values of Al—hTF are from references [53] [52]; the values of Nd—hTF and Sm—hTF are from reference [42]; the values of Lu—hTF and Gd—hTF are from reference [40]; the values of Tb—hTF (in bolder) are from reference [54]; the values of Sc—hTF are from reference[14]. All the data of metal*

*complexes of rare earth elements with* $CO_3^{2-}$ *and* $HCO_3^-$ *are from reference* [63]*. The data of other metal complexes of rare earth elements and* $CO_3^{2-}$ *are from reference* [64]*. The thermodynamic data for aqueous* $Ru^{3+}$ *is from reference* [65] *with large uncertainty (~ 4.7 kcal/mole). The free energy units are in kcal/mole. Data in parentheses are not used for regression analyses. The data for NTA is from reference* [66]*.*

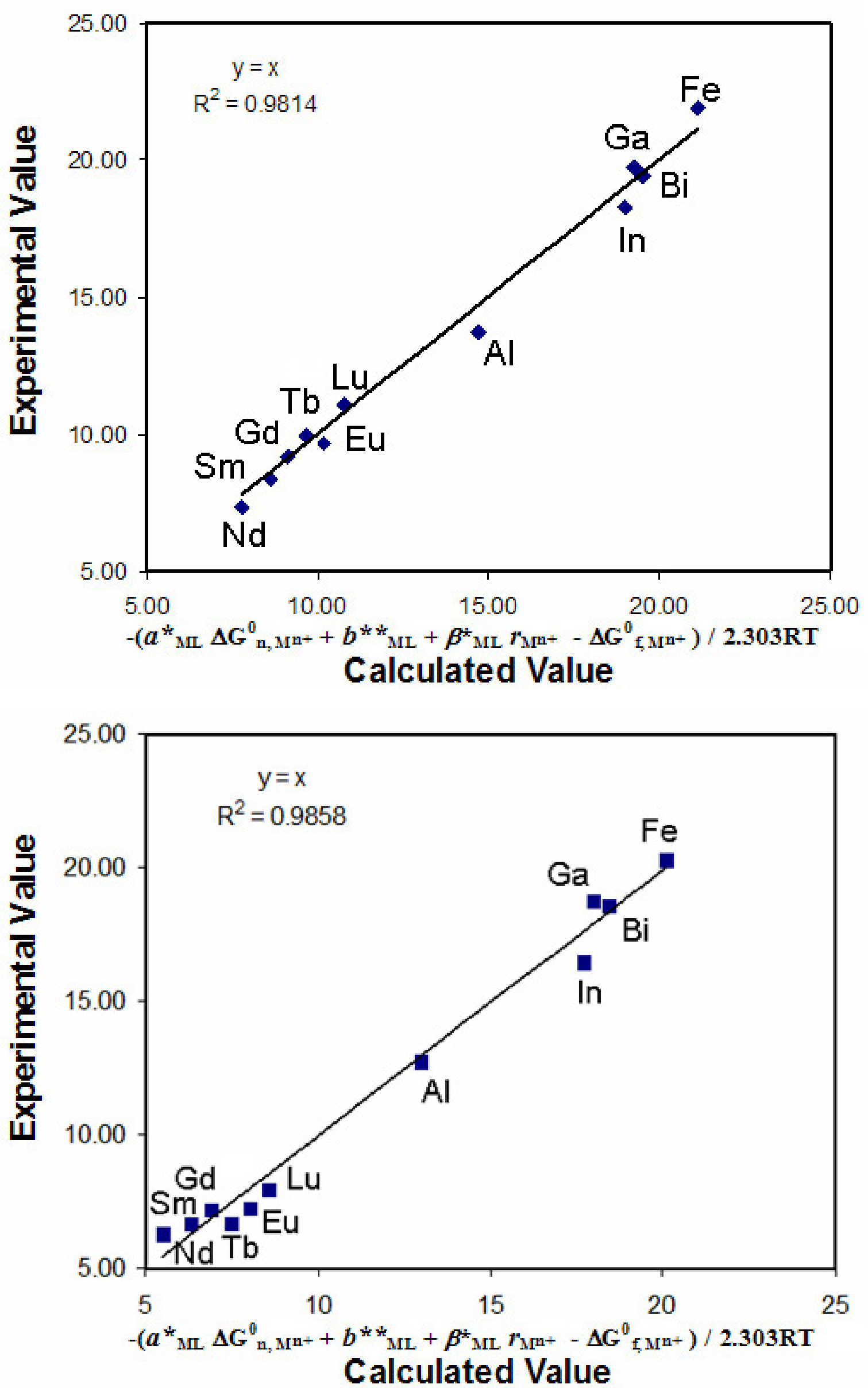

Figure 2: Linear free energy relationship of equation (1d) for $M^{3+}$—hTF (upper plot, $log K_1$) and $M^{3+}$— ($M^{3+}$- hTF) (lower plot, $log K_2$) complexes.

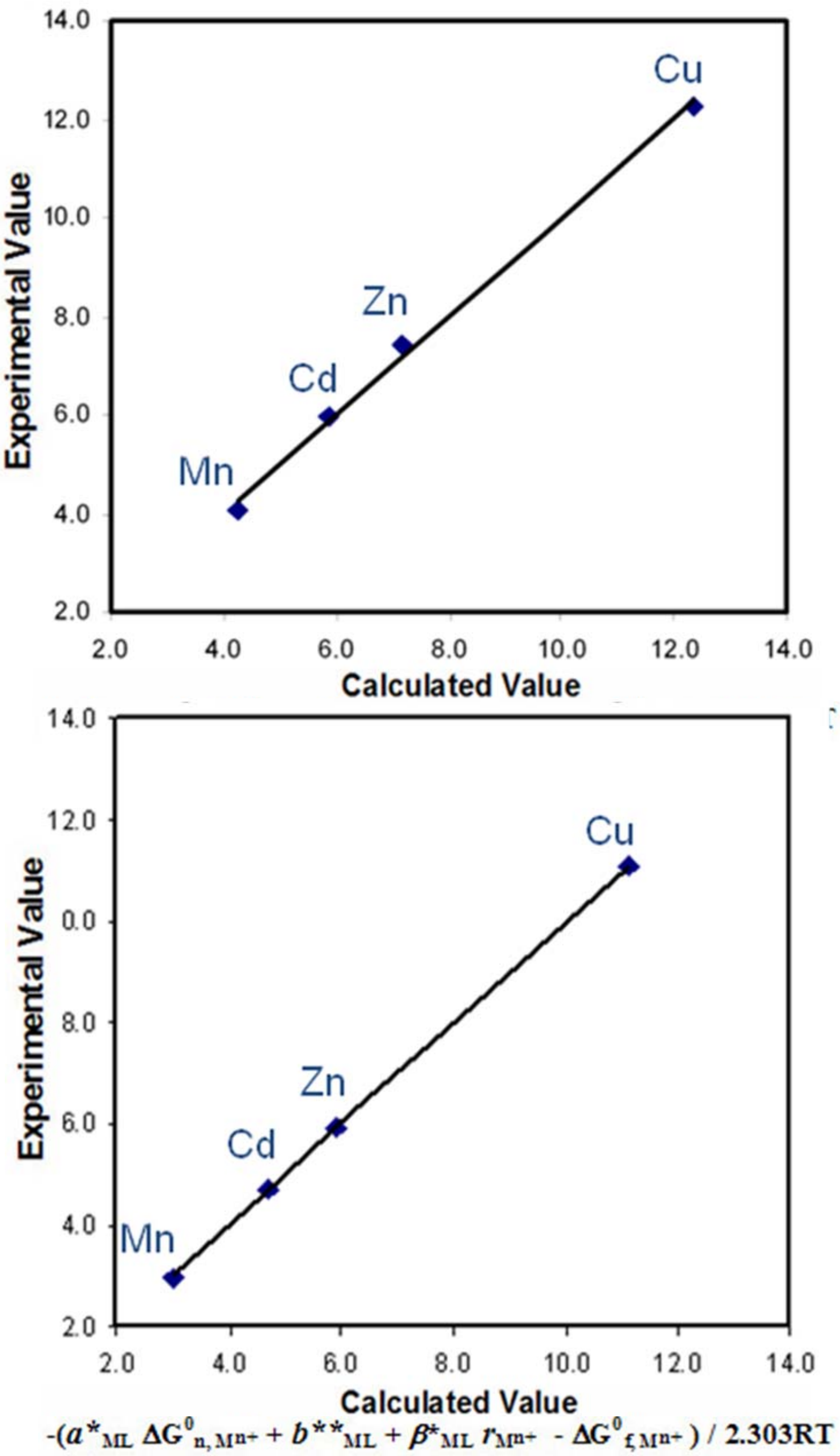

Figure 3: Linear free energy relationship of equation (1d) for $M^{2+}$—hTF (upper plot, *log* $K_1$) and $M^{2+}$— ($M^{2+}$- hTF) (lower plot, *log* $K_2$) complexes.

*Table 2: Ionic radii, thermodynamic data for divalent cations, and stability constants for metal—hTF, metal— carboxy-peptidase, and metal—ATP complex / chelate families.*

***Insert Table 2 Here***

*Note: Radii of the cations are from references [45, 49], effective radii of $NpO_2^{2+}$, $PuO_2^{2+}$, and $AmO_2^{2+}$ are assumed similar to that of $UO_2^{2+}$. The values of $\Delta G_f$ of the cations are from reference [45] except for $NpO_2^{2+}$ from reference [58], and $Pt^{2+}$, $Pd^{2+}$, $PuO_2^{2+}$ and $AmO_2^{2+}$ from reference [67]. The log K values of Mn—hTF, Zn—hTF, and Cd—hTF are from references [68 16 69]; the values of and Cu—hTF are from reference [70]; the values of and Ni—hTF are from reference [41]. The values of log K metal—ligand complexes are from reference [66]. The log K values of M — carboxy-peptidase are from reference[71].*

The equation (1c) also applies to $M^{2+}$—hTF ($log$ $K_1$) and $M^{2+}$—($M^{2+}$-hTF) ($log$ $K_2$) complexes as well (Figure 3, Table 2). There are two sets of data for Zn. We used the values selected by Harris [62] (Table 2). Standard errors (±0.16 and ±0.27) for the divalent cations are low due to limited experimental data. Only $Ni^{2+}$ data were not used for regression analysis, because of anomalously large discrepancies between the measured and calculated values. The predicted values of $log K_1$ and $log K_2$ for $Ni^{2+}$ are 9.6 and 7.9 respectively. By comparing the stability constants for $Cu^{2+}$ and $Co^{2+}$, it is expected that the predicted stability constants for $Ni^{2+}$ are between those for $Cu^{2+}$ and $Co^{2+}$ based on their positions in the Periodic table. It was explained that distortion of the octahedral site may result in the reduction of stabilization energy of the $Ni^{2+}$ [41]. We do not think this would cause such a large discrepancy. One possibility for the large difference between measured and calculated values is the coordination environment for Ni is different than for other divalent cations. This has been observed in divalent cations in calcite (C. N. = 6) and aragonite (C. N. = 9) polymorphs [47]. Future experimentation using different methods may be helpful to solve this problem. The equation also fits the experimentally determined stability constants of inorganic and organic metal complex families (Tables 1, 2, and 3), thus demonstrating the robustness of the proposed linear free energy relationship. Using the obtained $a^*_{ML}$, $b^{**}_{ML}$, and $\beta^*_{ML}$ values, the unknown stability constants metal—hTF complexes can be calculated (Table 1, Table 2). Given limited data availability for the divalent metals, the resulting predicted values should be considered as the first order approximation of the stability constants.

Our model shows that trivalent cations of $Au^{3+}$, $Tl^{3+}$, $Co^{3+}$, $V^{3+}$, $In^{3+}$, $Bi^{3+}$, $Cr^{3+}$, and $Mn^{3+}$ tend to compete with $Fe^{3+}$. The stability constants of $Tl^{3+}$, $Ru^{3+}$, $Au^{3+}$ and $Co^{3+}$ are higher than that of $Fe^{3+}$. It is reported that $Co^{3+}$—hTF complex is more stable than $Fe^{3+}$—hTF complex [72]. If $Co^{3+}$ is reduced to $Co^{2+}$, the stability constant of the $Co^{2+}$—hTF complex will be much lower than that of the $Co^{3+}$—hTF complex. The binding strengths for $Ru^{3+}$ ($log K_1 = 23.8$ and $log K_2 = 23.0$) are also predicted because $Ru^{3+}$-complexes were suggested as a potential anti-cancer agent [24 26 28 29 73]. It was suggested that compounds of Ru can be administered in $Ru^{3+}$ oxidation state (relatively inert), causing minimal damage to healthy cells, but being reduced to active in $Ru^{2+}$ oxidation state in cancer cells [74, 75]. The new binding constants may be used for calculating partitioning of between the $Ru^{3+}$-complexes and $Ru^{3+}$—hTF. Because of relatively strong binding between $Tl^{3+}$ and hTF and between $Co^{3+}$ and hTF, $Tl^{3+}$ and $Co^{3+}$ can compete with $Fe^{3+}$ in the $Fe^{3+}$—hTF complex. The predicted values of $log K_1$ and $log K_2$ for $Hg^{2+}$ are 11.1 and 10.3 respectively.

The linear free energy relationship can be used to predict the binding strengths of actinides to human serum transferring, which are difficult and dangerous to obtain experimentally [43, 76]. The predicted stability constants for $Cm^{3+}$—hTF and $Cm^{3+}$—(hTF-$Cm^{3+}$) complexes are 8.7 and 6.4 respectively. The predicted stability constants for $Am^{3+}$—hTF and $Am^{3+}$— (hTF-$Am^{3+}$) complexes are 8.2 and 6.0 respectively (Table 1). The values obtained here are larger than those predicted by Harris (1986) [41]. Using the obtained coefficients for divalent and trivalent cations (Table 3), we can also estimate the coefficients for tetravalent cations like U, Th, and Hf. There are very limited data for tetravalent cations. The behaviors of tetravalent cations may be different from those of trivalent cations [77]. Early results show that Ti may binds stronger that ferric Fe [78]. The coefficient $\beta^*_{ML}$ is proportional to the cation charge. Using available thermodynamic properties of tetravalent cations (Table 4), and the stability data for the Pu-hTF complex ($log\ K_1 = 21.25 \pm 0.75$) [43] and the Ti-hTF complex ($log\ K_1 = 26.8$) [79], it is possible to estimate the stability constants for other tetravalent cations if we can estimate values of $\beta^*_{ML}$ or $a^*_{ML}$. The value of $\beta^*_{ML}$ is related to ion charges based on regression values (Table 3). Based on $\beta^*_{ML}$ values for divalent cations and trivalent cations, we estimated a value of 150 kcal/mole (very similar to that for EDTA listed in Table 3) for the coefficients $\beta^*_{ML,}$ for tetravalent cations. We can calculate the coefficient of $a^*_{ML}$ to be 0.9563. The predicted stability constants for tetravalent cations are listed in table 4. The predicted stability constant for U-hTF complex is about 20. That is slightly lower than the value for the $Fe^{3+}$-hTF complex (Table 4). It is reported that the binding of the tetravalent cations (Ce, Hf, and Pu) could prevent normal closure of the transferrin interdomain cleft [80-82]. It is proposed that the cations with large hydration or solvation energies (like tetravalent cations) may attract water molecules at the binding sites, which affects the normal closure of the transferring domains [40].

*Table 3: Summary of regression coefficients for hTF and some metal complex families.*

***Insert Table 3 Here***

*Table 4: Table 4: Ionic radii, thermodynamic data for tetravalent cations, and predicted conditional stability constants of the $M^{4+}$-hTF complex.*

***Insert Table 4 Here***

*Note: Thermodynamic properties of tetravalent cations are from references [46, 50 51]. Gibbs free energies of formation ($\Delta G_f$) for $M^{4+}$-oxides are from references [58], [83], and [84] are also listed as an example. The predicted difference is small. Only stability data for Pu-hTF complex (21.25±0.75) [43] and Ti-hTF (26.8) [79] are available and used to constrain coefficient $b^{**}_{ML}$. Experimental data of Pu-hTF and Ti-hTF are in bold.*

With the calculated values, we can also estimate redox potentials for M-hTF with different oxidation states, such as, $Fe^{3+}$-hTF and $Fe^{2+}$-hTF using Nernst equation

$$E = 0.770 - 0.059 \log (K_c\text{-}Fe^{3+} / K_c\text{-}Fe^{2+}), \qquad (9)$$

Where 0.770V is the formal potential of the ferric to ferrous couple, and ($K_c$-$Fe^{3+}$ / $K_c$-$Fe^{2+}$ are the site-specific binding constants for ferric and ferrous ion, respectively [16].

As opposed to purely empirical methods, the linear free energy equation established above has a clear physical meaning for each term and thus provides a mechanistic basis for data extrapolation and interpolation. By substituting equation (2) into (1c), the following relationship can be obtained:

$$2.303RT\,log K_{ML} = (1 - a^*_{ML})\Delta G^0_{f,\,M^{n+}} - b^{**}_{ML} - \beta^*_{ML}\, r_{M^{n+}} + a^*_{ML}\,\Delta G^0_{s,\,M^{n+}}. \quad (10)$$

Interestingly, Gibbs free energies of formation of cations ($\Delta G^0_{f,\,M^{n+}}$) can be used as an index for the hardness/softness of metal cations (i.e., Lewis acids) proposed by Pearson [85, 86]. The more negative the value of the Gibbs free energy, the harder the acid will be. It is logical to postulate that the coefficient $a^*_{ML}$ or term 1-$a^*_{ML}$ characterizes the softness or hardness of complexing ligands (bases). A positive value of (1-$a^*_{ML}$) indicates a soft base or ligand, and a negative value of (1-$a^*_{ML}$) indicates a hard base or ligand. Therefore, the coefficient $a^*_{ML}$ can be used as an index of the hardness of ligands. Soft bases have values of $a^*_{ML} < 1$; and hard bases have values of $a^*_{ML} > 1$. Because of the correlation from the above equation, one is able to, as opposed to previous qualitative and or arbitrary numbering scales of hardness/softness, definitively express the hardness/softness based on the value of $a^*_{ML}$ relative to 1.

Overall, the stability constants are determined by both ionic radii of metal cations, hardness of cations (acids) and ligands (bases). The solvation energy ($\Delta G^0_{s,\,M^{n+}}$) term is also related to the radii of cations. Similarly, the coefficient $\beta^*_{ML}$ reflects the coordination environments of cations or the structural effects from metal-ligand binding. The term $\beta^*_{MXv} r_{M^{n+}}$ is similar to the steric effect; a large $\beta^*_{ML}$ value indicates a smaller coordination environment (polyhedron).

Previously proposed linear free energy relationships are based on M-hTF and metals with other ligands [5, 14], or they correlate *log*K values for one metal cation with *log*K values for another metal cation [41]. It is impossible to estimate *log*K for an element (like $Ru^{3+}$) without a known *log*K value with other ligand. Based on our linear free energy relationship, however, the *log*K values are determined by both the size and the energy ($\Delta G^0_{n,\ M^{n+}}$) of cations only. The previous methods may produce large uncertainties if they do not account for the effects from ionic radii and non-solvation energies. We can correlate the *log*K for M-hTF with *log*K for other metal-ligands using the obtained relationship. For instance, the difference between *log*K for hTF and *log*K for NTA (nitrilotriacetic acid) can be illustrated by using equation (1c):

$$logK_{hTF} = -(a^*_{hTF}\ \Delta G^0_{n,\ M^{3+}} + b^{**}_{hTF} + \beta^*_{hTF}\ r_{M^{3+}} - \Delta G^0_{f,\ M^{3+}}) / 2.303RT \qquad (11)$$

$$logK_{NTA} = -(a^*_{NTA}\ \Delta G^0_{n,\ M^{3+}} + b^{**}_{NTA} + \beta^*_{NTA}\ r_{M^{3+}} - \Delta G^0_{f,\ M^{3+}}) / 2.303RT \qquad (12)$$

$$2.303RT\ (logK_{NTA} - logK_{hTF}) = (a^*_{hTF} - a^*_{NTA})\ \Delta G^0_{n,\ M^{3+}} + (\beta^*_{hTF} - \beta^*_{NTA})\ r_{M^{3+}} + (b^{**}_{hTF} - b^{**}_{NTA}), \text{ or} \qquad (13)$$

$$2.303RT\ (logK_{NTA} - logK_{hTF}) = \Delta a^* \Delta G^0_{n,\ M^{3+}} + \Delta\beta^*\ r_{M^{3+}} + \Delta\ b^{**}. \qquad (14)$$

Using the coefficients in table 3, we can get

$$2.303RT\ (logK_{NTA} - logK_{hTF}) = -0.0312\Delta G^0_{n,\ M^{3+}} + 14.5\ r_{M^{3+}} - 7.44. \qquad (15)$$

The relationship can be schematically illustrated in Figure 4. For this particular case (see Table 3 for values of all coefficients), negative $\Delta a^*$ means NTA prefers hard acid (i.e., cations with large negative values of $\Delta G^0_{f,\ M^{n+}}$, such as rare earth elements and actinides); positive $\Delta\beta^*$ indicates that NTA (with lower $\beta^*$ value) prefers to bind large cations; Negative $\Delta\ b^{**}$ value means that hTF binds metals stronger than NTA does. However, in order to obtain a good correlation for the previous methods, we would need to find a ligand with same or close $\beta^*_{ML}$ without explicitly considering effects from radii of cations. The previous relationship overestimates *log*K values for large ions like rare earth elements because of the term $\Delta\beta^*\ r_{M^{n+}}$, and underestimate *log*K values for small ions, thus neglecting the importance of ionic radii.

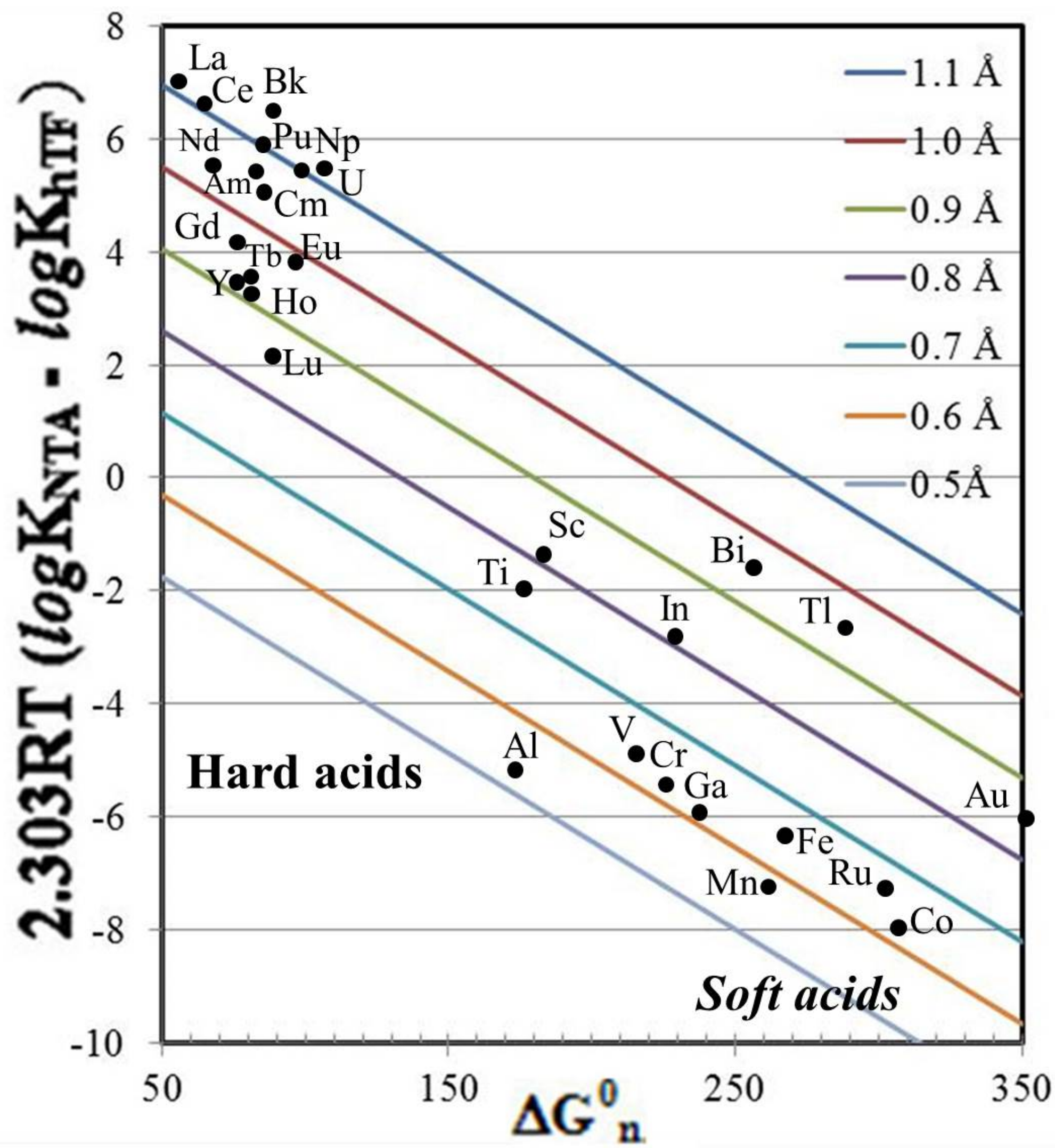

Figure 4: Diagram illustrates the binding strength relationship between M-hTF and M-NTA for trivalent cations. Both size and non-solvation energy of trivalent cations affect the difference between their stability constants.

The $Fe^{3+}$-hTF complex binds strongly to a receptor protein on the surface of cells. Inside the cell, the pH is lowering from 7.4 (extracellullar) to about 5.5 may cause the release of $Fe^{3+}$ from the $Fe^{3+}$—hTF complex [2,3,6,10,11,13]. There are four proposed mechanisms for the release of $Fe^{3+}$ inside the cell [1,2]. The four are acidification, ligand exchange, reduction of $Fe^{3+}$, and synergistic anion exchange [2]. The reduction mechanism explains Fe well [87,88], but not for the non-redox elements. A recent molecular dynamic modeling result indicates that the pH dependent change in the dynamics is traced to the altered electrostatic potential distribution along the surface [89]. It was proposed that protonation of carbonate [10], protonation of His 249 [11], and both carbonate and His 249 ligand protonation [6] are key steps in iron release. According to our linear free energy relationship for ligands and protonated ligands, we propose that the protonation of a ligand (e.g., ATP, and carbonate) can result in hardness increasing (i.e., larger $a^*_{ML}$ value) besides overall binding strength decrease (Table 3), which can lower the stability constants of soft acids (i.e., cations with high Gibbs free energy of formation in here) dramatically[85].

We can apply this concept to protonated hTF (or the formation of the H-hTF protein ligand) in acidic inside cell environment, which can lower the stability constant (*log* K) of the $M^{3+}$—H-hTF complexes for the soft acids like $Fe^{3+}$ dramatically and result in $M^{3+}$ being released from the $M^{3+}$—H-hTF complex. The protonation of ligand hTF is similar to the mechanism of acidification. The protonation of a ligand decreases stability of protonated metal complexes. For instance, in basic environment, $CO_2$ in water is dominated by $CO_3^{2-}$ ligand. As the pH of a solution decreases, the $CO_3^{2-}$ ligand becomes protonated (or, ligand $HCO_3^-$ will form). The protonated carbonate $HCO_3^-$ ligand is harder than the unprotonated $CO_3^{2-}$ ligand (i.e., larger $a^*_{ML}$ value with respect to that of carbonate ligand, Table 3). We propose that the overall function of hTF ligand in basic and acidic solutions is similar to the protonation of $CO_3^{2-}$ ligand, even though the function and property of the synergistic anion $CO_3^{2-}$ is not the same as that of isolated $CO_3^{2-}$. Such kind of a transformation mechanism may be useful for designing engineered proteins that can uptake toxic metals in a relatively basic solution and release the coordinated metals in a relatively acidic solution.

## Conclusions

The linear free energy relationship developed here can be used to predict unknown stability constants of metal—hTF protein complexes from limited experimental data. The discrepancies between the predicted and experimental data are generally less than 0.6 *log* units, far better than other empirical methods. The stability constants for actinides (e.g., Pu and Am) and tetravalent cations are also predicted. This relationship can be applied to the binding between metal cations and enzymes, and between metal metals and engineered proteins reported by Vita et al. [90]. It can also be used to systematically evaluate the quality of data collected from multiple sources and to select internally consistent and reliable data sets for metal speciation in physiological fluid. It is also proposed that the release of $Fe^{3+}$ inside cells results from protonation of the protein hTF and formation of the harder ligand H-hTF with a large $a^*_{ML}$ value that reduces the binding strength of soft Lewis acids like $Fe^{3+}$ dramatically. The proposed method can be used as a guideline for designing engineered proteins and compounds with desired selectivity and strength [91, 92 93].

**Acknowledgment:** Authors thank David C. Xu for helping to organize some data. This work is supported by the NASA Astrobiology Institute under grant N07-5489.

## Notes and References

Table 1: Ionic radii, thermodynamic data for trivalent cations, and stability constants for three families of metal complexes.

| $M^{3+}$ | $r_{M^{3+}}$ (Å) | $\Delta G_s$ $M^{3+}_{(aq)}$ | $\Delta G_f$ $M^{3+}_{(aq)}$ | $\Delta G_n$ $M^{3+}_{(aq)}$ | *log $K_{ML}$* | | | | | | | |
|---|---|---|---|---|---|---|---|---|---|---|---|---|
| | | | | | M—hTF (Exper.) | M—hTF (Calc. ±0.54) | M—(hTF-M) (Exper.) | M—(hTF-M) (Calc. ±0.76) | $(M—CO_3)^+$ (Exper.) | $(M—CO_3)^+$ (Calc. ±0.24) | M-NTA (Exper.) | M-NTA (Calc. ±0.36) |
| Al | 0.53 | -288.10 | -115.38 | 172.73 | 12.4; **13.72** | 14.63 | 11.8; **12.72** | 12.98 | 8.43 | 8.41 | | 10.49 |
| Cr | 0.62 | -276.92 | -51.50 | 225.42 | | 18.40 | | 17.12 | | 8.97 | | 13.91 |
| Fe | 0.65 | -273.11 | -4.12 | 268.99 | 21.90 | 21.06 | 20.60 | 20.08 | 9.72 | 9.38 | 15.90 | 15.87 |
| Co | 0.63 | -275.01 | 32.03 | 307.04 | | 23.16 | | 22.43 | | 9.71 | | 16.94 |
| Ga | 0.62 | -276.28 | -38.00 | 238.28 | 19.75 | 19.17 | 18.80 | 17.98 | 8.72 | 9.09 | 13.80 | 14.43 |
| Ti | 0.76 | -259.08 | -83.60 | 175.48 | | 15.96 | | 14.34 | | 8.52 | | 14.05 |
| V | 0.64 | -273.74 | -57.90 | 215.84 | | 18.00 | | 16.66 | | 8.90 | | 13.98 |
| Mn | 0.58 | -281.46 | -20.30 | 261.16 | | 20.18 | | 19.14 | | 9.26 | | 14.52 |
| Sc | 0.81 | -253.25 | -140.20 | 113.05 | (14.60) | 12.30 | (13.30) | 10.25 | | 7.91 | 12.70 | 12.31 |
| Tl | 0.95 | -237.77 | 51.30 | 289.07 | | 21.46 | | 20.62 | | 9.27 | | 18.84 |
| Bi | 0.96 | -236.71 | 19.79 | 256.50 | 19.42 | 19.50 | 18.58 | 18.45 | | 8.96 | 18.20 | 17.73 |
| In | 0.81 | -253.25 | -23.40 | 229.85 | 18.30 | 18.96 | 16.60 | 17.72 | | 8.96 | | 16.31 |
| Y | 0.92 | -240.99 | -163.80 | 77.19 | | 9.62 | | 7.34 | 6.94 | 7.43 | 11.47 | 11.56 |
| La | 1.14 | -218.51 | -164.00 | 54.51 | | 5.58 | | 3.14 | 6.82 | 6.64 | 10.47 | 10.22 |
| Ce | 1.07 | -225.39 | -161.60 | 63.79 | | 7.19 | | 4.81 | 6.95 | 6.94 | 10.70 | 10.92 |
| Pr | 1.06 | -226.39 | -162.60 | 63.79 | | 7.33 | | 4.95 | 7.03 | 6.97 | 10.87 | 10.96 |
| Nd | 1.04 | -228.41 | -160.60 | 67.81 | 7.33 | 7.82 | 6.28 | 5.47 | 7.13 | 7.06 | 11.10 | 11.17 |
| Pm | 1.06 | -226.39 | -158.00 | 68.39 | | 7.59 | | 5.24 | 7.22 | 7.01 | | 11.12 |
| Sm | 1.00 | -232.52 | -159.10 | 73.42 | 8.37 | 8.63 | 6.63 | 6.32 | 7.30 | 7.22 | 11.32 | 11.45 |
| Eu | 0.98 | -234.60 | -137.30 | 97.30 | 9.66 | 10.21 | 7.27 | 8.06 | 7.37 | 7.48 | 11.32 | 12.28 |
| Gd | 0.97 | -235.65 | -158.60 | 77.05 | 9.20 | 9.16 | 7.18 | 6.87 | 7.44 | 7.32 | 11.35 | 11.59 |
| Tb | 0.93 | -239.91 | -159.50 | 80.41 | 11.2;10.96; **9.96** | 9.73 | 8.52; 7.61; **6.34** | 7.46 | 7.50 | 7.44 | 11.50 | 11.68 |
| Dy | 0.92 | -240.99 | -158.70 | 82.29 | | 9.92 | | 7.66 | 7.55 | 7.47 | 11.63 | 11.73 |
| Ho | 0.91 | -242.08 | -161.40 | 80.68 | | 9.90 | | 7.64 | 7.59 | 7.48 | 11.76 | 11.66 |
| Er | 0.89 | -244.26 | -159.90 | 84.36 | | 10.26 | | 8.02 | 7.63 | 7.55 | 11.90 | 11.73 |
| Tm | 0.87 | -246.59 | -159.90 | 86.69 | | 10.52 | | 8.30 | 7.66 | 7.60 | 12.07 | 11.73 |
| Yb | 0.86 | -247.81 | -153.00 | 94.81 | | 11.05 | | 8.88 | 7.67 | 7.69 | 12.21 | 11.96 |
| Lu | 0.85 | -248.94 | -159.40 | 89.54 | 11.08 | 10.80 | 7.93 | 8.59 | 7.70 | 7.66 | 12.32 | 11.73 |
| U | 1.12 | -220.45 | -113.88 | 106.57 | | 8.88 | | 6.80 | | 7.17 | | 12.13 |
| Pu | 1.08 | -224.39 | -138.15 | 86.24 | | 8.33 | | 6.11 | | 7.12 | | 11.65 |
| Np | 1.10 | -222.41 | -123.59 | 98.82 | | 8.75 | | 6.61 | | 7.17 | | 11.98 |
| Am | 1.07 | -225.39 | -143.19 | 82.20 | | 8.24 | | 5.99 | (6.5) | 7.11 | 11.50 | 11.55 |
| Ac | 1.20 | -212.80 | -152.96 | 59.84 | | 4.83 | | 2.42 | | 6.47 | | 9.94 |
| Cm | 1.05 | -227.40 | -142.40 | 85.00 | | 8.67 | | 6.44 | | 7.19 | 11.80 | 11.72 |
| Bk | 1.04 | -228.41 | -138.86 | 89.55 | | 9.07 | | 6.86 | | 7.26 | | 11.91 |
| Au | 0.85 | -248.94 | 103.60 | 352.31 | | 25.79 | | 25.38 | | 10.01 | | 20.73 |
| Ru | 0.67 | -269.98 | 41.44 | 311.42 | | 23.60 | | 22.91 | | 9.76 | | 17.68 |

hTF = human serum transferring; NTA = Nitrilotriacetic acid ($C_6H_9O_6N$)

Note: Radii of the trivalent cations are from reference [50]. The values of $\Delta G_f$ of the cations are from references [51, 55] except for $Bi^{3+}$ from reference [56], $Ti^{3+}$ from reference [57] , and $Pu^{3+}$, $Np^{3+}$, and $Am^{3+}$ from references [58 59]. The *log* K values of Fe—hTF are from references [13]; the values of Bi—hTF, Ga—hTF, and In—hTF are from references [15, 60 61 62]; the values of Al—hTF are from references [53 52]; the values of Nd—hTF and Sm—hTF are from reference [42]; the values of Lu—hTF and Gd—hTF are from reference [40]; the values of Tb—hTF (in bolder) are from reference [54]; the values of Sc—hTF are from reference[14]. All the data of metal complexes of rare earth elements with $CO_3^{2-}$ and $HCO_3^-$ are from reference [63]. The data of other metal complexes of rare earth elements and $CO_3^{2-}$ are from reference [64]. The thermodynamic data for aqueous $Ru^{3+}$ is from reference [65] with large uncertainty (~ 4.7 kcal/mole). The free energy units are in kcal/mole. Data in parentheses are not used for regression analyses. The data for NTA are from reference [66].

Table 2 Ionic radii, thermodynamic data for divalent cations, and stability constants for metal—hTF, metal— carboxy-peptidase, and metal—ATP complex / chelate families.

| | | | | | | log $K_{ML}$ | | | | | | | | |
|---|---|---|---|---|---|---|---|---|---|---|---|---|---|---|
| $M^{2+}$ | $r_M^{2+}$ (Å) | $\Delta G_s$ $M^{2+}_{(aq)}$ | $\Delta G_f$ $M^{2+}_{(aq)}$ | $\Delta G_n$ $M^{2+}_{(aq)}$ | M—hTF (Exper.) | M—hTF (Calc. ±0.16) | M—(M-hTF) (Exper.) | M—(M-hTF) (Calc. ±0.27) | M—ATP (Exper.) | M—ATP (Calc. ±0.28) | M—(H-ATP) (Exper.) | M—(H-ATP) (Calc. ±0.19) | M-carboxy-peptidase (Exper.) | M-carboxy-peptidase (Calc. ±0.94) |
| Be | 0.45 | -175.02 | -89.80 | 85.22 | | 2.29 | | 0.86 | 4.06 | 3.93 | 4.55 | 4.69 | | -11.86 |
| Mg | 0.72 | -145.80 | -108.83 | 36.97 | | -0.03 | | -1.37 | 4.76 | 4.72 | 4.14 | 4.27 | | -5.34 |
| Ca | 1.00 | -121.28 | -132.12 | -10.84 | | -6.45 | | -7.70 | 4.85 | 4.91 | 4.35 | 4.18 | | -2.52 |
| Mn | 0.82 | -136.46 | -55.20 | 81.26 | 4.06 | 4.26 | 2.96 | 3.00 | 3.77 | 3.76 | 4.69 | 4.79 | 4.60 | 4.20 |
| Fe | 0.77 | -141.04 | -21.87 | 119.17 | | 8.20 | | 6.94 | | 5.21 | | 4.10 | | 7.42 |
| Co | 0.74 | -144.35 | -13.00 | 131.35 | | 9.45 | | 8.18 | 4.63 | 5.17 | 4.19 | 4.05 | 7.00 | 7.71 |
| Ni | 0.70 | -147.75 | -10.90 | 136.85 | (4.10) | 9.96 | (3.23) | 8.68 | 5.02 | 5.14 | 4.23 | 4.06 | 8.20 | 7.05 |
| Cu | 0.73 | -144.83 | 15.55 | 160.38 | 12.26 | 12.36 | 11.08 | 10.11 | 6.13; **6.0** | 5.52 | 3.52 | 3.86 | 10.60 | 11.35 |
| Zn | 0.75 | -143.30 | -35.17 | 108.13 | 7.42 | 7.19 | 5.91 | 5.91 | | 3.49 | | 5.04 | (10.50) | 5.10 |
| Sr | 1.16 | -109.30 | -133.72 | -24.42 | | -10.24 | | -11.42 | | 5.10 | | 4.08 | | -0.59 |
| Cd | 0.95 | -125.31 | -18.57 | 106.74 | 5.95 | 5.89 | 4.86 | 4.70 | 3.54 | 3.61 | 5.04 | 4.92 | 10.80 | 11.63 |
| Sn | 1.11 | -112.91 | -6.63 | 106.28 | | 3.73 | | 2.61 | | 5.22 | | 4.02 | | 15.60 |
| Ba | 1.36 | -95.99 | -132.73 | -36.74 | | -15.72 | | -16.82 | | 5.28 | | 4.03 | | 1.22 |
| Eu | 1.17 | -108.59 | -129.10 | -20.51 | | -10.02 | | -11.20 | 3.29 | 3.29 | 5.09 | 5.19 | | 0.13 |
| Hg | 1.02 | -119.71 | 39.36 | 159.07 | | 10.34 | | 9.21 | | 3.66 | | 4.90 | 21.00 | 20.43 |
| Pb | 1.18 | -107.89 | -5.79 | 102.10 | | 2.09 | | 1.00 | | 5.93 | | 3.67 | | 16.52 |
| Ra | 1.39 | -94.14 | -134.20 | -40.06 | | -16.80 | | -17.89 | | 3.21 | | 5.25 | | 1.19 |
| $UO_2$ | 0.754 | -142.54 | -227.70 | -85.16 | | -12.27 | | -13.68 | | 2.47 | | 5.45 | | -20.15 |
| $NpO_2$ | 0.73 | -148.60 | -190.20 | -41.60 | | -10.67 | | -12.05 | | 2.53 | | 5.50 | | -18.09 |
| $PuO_2$ | 0.71 | -150.59 | -183.50 | -32.91 | | -9.85 | | -11.24 | | 2.59 | | 5.48 | | -17.77 |
| $AmO_2$ | 0.70 | -151.60 | -156.70 | -5.10 | | -7.10 | | -8.48 | | 2.91 | | 5.32 | | -14.52 |
| Pd | 0.80 | -141.87 | 42.49 | 184.36 | | 12.03 | | 10.83 | | 5.58 | | 3.91 | | 14.48 |
| Pt | 0.80 | -141.87 | 54.80 | 196.67 | | 13.27 | | 10.98 | | 5.73 | | 3.83 | | 16.10 |

Note: Radii of the cations are from references [45, 49], effective radii of $NpO_2^{2+}$, $PuO_2^{2+}$, and $AmO_2^{2+}$ are assumed similar to that of $UO_2^{2+}$. The values of $\Delta G_f$ of the cations are from reference [45] except for $NpO_2^{2+}$ from references [58], and $Pt^{2+}$, $Pd^{2+}$, $PuO_2^{2+}$ and $AmO_2^{2+}$ from reference [67]. The *log* K values of Mn—hTF,

Zn—hTF, and Cd—hTF are from reference [68] [16] [69]; the values of and Cu—hTF are from reference [70]; the values of and Ni—hTF are from reference [41];. The values of *log* K metal—ligand complexes are from reference [66]. The *log* K values of M — carboxy-peptidase are from reference [71]. Values in parentheses are not used for regression analysis due to large difference between experimental and calculated values. The large difference for Zn-carboxy-peptidase between measured and calculated values may tell that the coordination environment for Zn is different from other divalent cations.

Table 3: Summary of regression coefficients for hTF and some metal complex families.

| Metal complex type | Regression coefficients | | | |
|---|---|---|---|---|
| $M^{n+}L$ | $a^*_{ML}$ | $b^{**}_{ML}$ | $\beta^*_{ML}$ | Data points |
| | **Divalent Cations** | | | |
| $M^{2+}$-Carboxypeptidase | 0.8199 | -162.89 | 43.1 | 6 |
| $M^{2+}$—hTF | 0.8632 | -209.46 | 95.5 | 4 |
| $M^{2+}$—(hTF-$M^{2+}$) | 0.8623 | -207.17 | 94.9 | 4 |
| $M^{2+}$—EDTA * | 0.9046 | -207.33 | 71.9 | 12 |
| $M^{2+}$—ATP# | 0.9749 | -27.96 | 8.4 | 9 |
| $M^{2+}$—(H-ATP)# | 1.0090 | +15.79 | -8.7 | 9 |
| | **Trivalent Cations** | | | |
| $M^{3+}$—hTF | 0.9221 | -357.79 | 119.2 | 11 |
| $M^{3+}$—(hTF-$M^{3+}$) | 0.9128 | -353.92 | 119.2 | 11 |
| $M^{3+}$—EDTA * | 0.9284 | -350.35 | 107.8 | 26 |
| $M^{3+}$—NTA * | 0.9533 | -350.35 | 105.7 | 22 |
| $M^{3+}$—$CO_3^{2+}$ | 0.9894 | -358.57 | 117.0 | 18 |
| $M^{3+}$—$HCO_3^{2+}$ | 1.0300 | -345.41 | 108.0 | 15 |
| | **Tetravalent Cations** | | | |
| $M^{4+}$—EDTA * | 0.9688 | -527.42 | 153.5 | 4 |
| $M^{4+}$—hTF | 0.9563 | -514.33 | 150.0 | |

Note: All values refer to standard condition (25°C, 1 bar).

hTF = human serum transferring; NTA = Nitrilotriacetic acid ($C_6H_9O_6N$)

ATP = Adenosine 5'-triphosphate; H-ATP = protonated ATP

* Use the available data for ionic strength of 0.1 for metal-EDTA and $M^{3+}$—NTA chelates.

# Aqueous $M^{2+}$—ligand mono-dendate complexes require additional term from solvation energy

($c_{ML} \bullet \Delta G^0_{s, M^{2+}}$) because of strong effect from solvent water, i.e.,

$\Delta G^0_{f, ML} = a_{ML}\, \Delta G^0_{n, M^{2+}} + c_{ML}\, \Delta G^0_{s, M^{2+}} + \beta^*_{ML}\, r_{M^{2+}} + b_{ML}$, or

$2.303RT\,log K_{ML} = (1 - a_{ML})\Delta G^0_{f, M^{n+}} - \beta^*_{ML}\, r_{M^{n+}} + (a_{ML} - c_{ML})\Delta G^0_{s, M^{n+}} - b^{**}_{ML}$.

The $c_{ML}$ values for $M^{2+}$—ATP and $M^{2+}$—AH-TP are 0.9576 and 1.0180 respectively

(see Xu et al., 2017)[85] for details.

Table 4: Ionic radii, thermodynamic data for tetravalent cations, and predicted conditional stability constants of $M^{4+}$-hTF complex

| | | | | | $\Delta G_f$ | | calculated *log* $K_1$ |
|---|---|---|---|---|---|---|---|
| $M^{4+}$ | $r_{M^{4+}}$ (Å) | $\Delta G_s$ $M^{4+}_{(aq)}$ | $\Delta G_f$ $M^{4+}_{(aq)}$ | $\Delta G_n$ $M^{4+}_{(aq)}$ | $MO_2$ (experimental) | $MO_2$ (calculated) | M-hTF ( $\beta^*$ = 150) |
| Zr | 0.79 | -373.11 | -141.00 | 232.11 | -248.50 | -249.26 | 23.44 |
| Hf | 0.78 | -374.41 | -156.80 | 217.61 | -260.09 | -259.32 | 23.13 |
| Ce | 0.94 | -354.23 | -120.44 | 233.79 | -244.40 | -243.32 | 20.83 |
| Th | 1.02 | -344.65 | -168.52 | 176.13 | -279.34 | -279.50 | 17.24 |
| U | 0.97 | -350.60 | -124.40 | 226.20 | -246.62 | -247.46 | 19.96 |
| Np | 0.95 | -353.02 | -120.20 | 232.82 | -244.22 | -243.66 | 20.59 |
| Pu | 0.93 | -355.45 | -114.96 | 240.49 | -238.53 | -239.14 | **21.25** |
| Am | 0.92 | -356.68 | -89.20 | 267.48 | (-230; -210.4) | -221.34 | 22.30 |
| Po | 1.06 | -339.98 | 70.00 | 409.98 | | -121.13 | 23.65 |
| Pb | 0.84 | -366.68 | 143.50 | 510.18 | | -60.87 | 31.44 |
| Sn | 0.71 | -383.69 | 40.50 | 424.19 | | -122.79 | 30.55 |
| Ti | 0.68 | -387.76 | -93.00 | 294.76 | | -210.69 | **26.78** |
| Mn | 0.60 | -398.88 | 58.00 | 456.88 | | -104.35 | 32.54 |

Note: Thermodynamic properties of tetravalent cations are from references [46, 50 51]. Gibbs free energies of formation ($\Delta G_f$) for $M^{4+}$-oxides are from references [58], [83], and [84] are also listed as an example. The predicted difference is small. Only stability data for Pu-hTF complex (21.25±0.75) [43] and Ti-hTF (26.8) [79] are available and used to constrain coefficient $b^{**}_{ML}$. Experimental data of $Pu^{4+}$-hTF and $Ti^{4+}$-hTF are in bold cases.